\documentclass[preprint2]{aastex63}

\usepackage{CJK}
\usepackage{amsmath}
\usepackage{bm}
\usepackage{multirow}

\newcommand{\petit}{\texttt{petitRADTRANS}}

\newcommand{\pmn}{\texttt{PyMultiNest}}

\received{\today}
\revised{}
\accepted{}
\submitjournal{ApJ}

\shorttitle{HIP 65426}
\shortauthors{Wang}
\graphicspath{{./}{figures/}}

\begin{document}
\begin{CJK*}{UTF8}{gbsn}

\title{Spectral Retrieval with JWST Photometric data: a Case Study for HIP 65426~\lowercase{b} }

\correspondingauthor{Ji Wang}
\email{wang.12220@osu.edu}

\author[0000-0002-4361-8885]{Ji Wang (王吉)}
\affiliation{Department of Astronomy, The Ohio State University, 100 W 18th Ave, Columbus, OH 43210 USA}

\begin{abstract}

Half of the JWST high-contrast imaging objects will only have photometric data {{as of Cycle 2}}. However, to better understand their atmospheric chemistry which informs formation origin, spectroscopic data are preferred. Using HIP 65426 b, we investigate to what extent planet properties and atmospheric chemical abundance can be retrieved with only JWST photometric data points (2.5-15.5 $\mu$m) in conjunction with ground-based archival low-resolution spectral data (1.0-2.3 $\mu$m). {{We find that the data is consistent with an atmosphere with solar metallicity and C/O ratios at 0.40 and 0.55. We rule out 10x solar metallicity and an atmosphere with C/O = 1.0. }}We also find strong evidence of silicate clouds but no sign of an enshrouding featureless {{dust}} extinction. This work offers guidance and cautionary tales on analyzing data in the absence of medium-to-high resolution {{spectral}} data. 

\end{abstract}



\section{Introduction}
\label{sec:intro}

Directly-imaged exoplanets provide a unique window to understand atmospheric structure and chemistry. This information shed light on the formation pathway and evolution history. {{For example, it is still being debated if planetary mass objects $<$13 M$_{\rm{Jupitr}}$) at wide separation ($\gtrapprox$10 AU) form similarly to or differently from their higher-mass counterparts brown dwarfs~\citep[BDs,][]{Bonnefoy2018,Wang2020}.}} Studying their atmospheric chemical composition can offer important clues~\citep{Oberg2011}. 

Chemical composition is best studied with spectroscopy whereby spectral lines {{or bands}} are resolved and chemical abundances can be measured. {{Conventionally, spectroscopic data come from ground-based low-to-medium resolution (R$<$5000) integral field unit~\citep[IFU, e.g.,][]{Larkin2006,Groff2015,Macintosh2014,Beuzit2019}. In the coming years, our understanding of the formation and evolution of directly-imaged exoplanets will be revolutionized thanks to the space-based facility JWST. Indeed, JWST Early Release Science (ERS) program~\citep{Carter2022,Miles2023} has given us a glimpse of the opportunities and challenges in modeling and interpreting atmospheres with JWST data.}}



One challenge for JWST data is that not every object will receive spectroscopic observations. {{Half of the JWST high-contrast imaging objects will only have photometric measurements with NIRCam and MIRI through Cycle-2. While some of them will have spectroscopic data in future cycles, certain objects---especially those too faint or too embedded---are too challenging to have spectroscopic data.}} This work attempts to address to what extent one can characterize planetary atmospheres with JWST photometric data points in conjunction with archival ground-based low-spectral resolution data.  

{{We use HIP 65426 b---a JWST target in the Early Release Science (ERS) program~\citep{Hinkley2022}---in this case study. The ERS data for HIP 65426 b~\citep{Carter2022} including photometric data from NIRCam~\citep{Rieke2005} and MIRI~\citep{Rieke2015} that cover a wavelength range from $\sim$2 to 16 $\mu$m.}} This paper is organized as follows. We briefly introduce our retrieval framework in \S \ref{sec:method}. Lessons learned from testing the retrieval framework on mock data are given in \S \ref{sec:test}. Main results on retrieving HIP 65426 b properties are in \S \ref{sec:results}. Discussions are provided in \S \ref{sec:discussion}. A summary of the paper can be found in \S \ref{sec:summary}.

\section{Retrieval Framework}
\label{sec:method}

We refer to~\citet{Wang2020, Wang2022, Wang2023} for details of the retrieval framework with which we perform retrieval analyses. In summary, we model exoplanet atmospheres based on \petit\ and consider both low and high resolution modes (R=1,000 and R=1,000,000) {{when such data are available}}. For the temperature profile, we adopt a flexible P-T profile as described in~\citet{Petit2020}. To sample the posterior distribution in a Bayesian framework, we used \pmn~\citep{Buchner2014}. 

We include MgSiO$_3$ clouds~\citep{Molliere2020} with a new addition of featureless extinction. The extinction $\tau$ follows the exponential relation with wavelength ($\lambda$) such that $\tau(\lambda) = \exp(-\alpha\cdot{\frac{1\ \mu m}{\lambda}})$~\citep{Gordon2003}, {{where $\alpha$ is the extinction coefficient}}. Adding the wavelength-dependent but spectrally featureless extinction is motivated by the inferred circum-planetary dust surrounding PDS 70 planets~\citep{Wang2021b}. We would like to investigate if circum-planetary dust is required to explain the spectral energy distribution of HIP 65426 b. A full list of parameters and their priors are in Table \ref{tab:prior}. 

\section{Testing With Mock Data}
\label{sec:test}
\subsection{Generating Mock Data}
\label{sec:mock}
 
Our mock data are resampled modeled spectra from \petit. To generate model spectra, we use the model parameters listed in {{the ``Input" column of}} Table \ref{tab:mock_result}. Four cases are considered in terms of metallicity (1x and 10x solar) and the mixing ratio of MgSiO$_3$ (low at -4 dex and high at -3 dex). When calculating the corresponding mixing ratio of CO, H$_2$O, CH$_4$, and CO$_2$, we use \texttt{poor\_mans\_nonequ\_chem} to interpolate a pre-calculated chemical grid from \texttt{easyCHEM}~\citep{Molliere2017}. We consider a quench case in which mixing ratios are homogenized above 10 bar and set {{by}} the chemistry at 10 bar and 2500 K. The {{input}} values of parameters in Table \ref{tab:mock_result} may not be physically plausible, e.g., 4.0 for log(g) and 3.5 for planet radius, but the choices are based on the metric that the emerging fluxes are roughly consistent between the mock and actual data.  

To simulate the mock data, we resample a model spectrum to the wavelength grid of existing data. We use the same data set that is used in~\citet{Carter2022}. Table \ref{tab:data} tabulates the data set including VLT/SPHERE-IFS between 1.00 and 1.65 μm~\citep{Chauvin2017},
VLT/SPHERE-IRDIS $H$ and $K$-band photometry~\citep{Cheetham2019}, and JWST NIRCam and MIRI
photometry~\citep{Carter2022}. Since the spectral resolution of low-resolution data is not uniform across the wavelength range, sampling the synthetic spectrum to the wavelength grid of existing data ensures that the synthetic spectrum has the same varying spectral resolution as the original spectrum. {{For JWST photometric points, we convolve the response profile of each filter~\citep{Rodrigo2020} with the model spectrum to compare to the data.}}

\subsection{Analyzing Mock Data}
\label{sec:analyze_mock}

We would like to answer the following major questions through the exercise of analyzing mock data: (1) can we measure the extinction coefficient $\alpha$; (2), can we constrain the mixing ratio of the MgSiO$_3$ cloud; (3) can we constrain the the mixing ratios for molecular species given the data quality?

We start with a more constraining condition in which we fix chemical abundance to 1x or 10x solar and cloud properties (as shown in Table \ref{tab:mock_result}). The retrieved value for  extinction coefficient $\alpha$ is usually within 2-$\sigma$ to the input value of 3.0. The retrieval uncertainty is typically 0.1-0.3 dex with the uncertainty on the higher end for 10x solar metallicity and a higher mixing ratio of MgSiO$_3$ at -3 dex. The higher uncertainty is due to the lower flux because of the higher extinction of the MgSiO$_3$ cloud and the higher opacity of molecules. The flux for this case is $\sim$20 lower than the case with 1x solar metallicity and  {{a lower}} mixing ratio of MgSiO$_3$ at -4 dex. 

We then relax some constraints to allow for the variation of mixing ratios for molecules and the cloud species MgSiO$_3$. We still fix other cloud parameters. Furthermore, we also fix planet radius at the input value of 3.5 R$_{\rm{Jupiter}}$ to limit the covariance between planet radius and surface gravity. After performing a retrieval analysis on the data with 1x solar metallicity, we find that the retrieved $\alpha$ is consistent with the input value with an uncertainty of $\sim$0.2 dex. The mixing ratio of MgSiO$_3$ is retrieved at a lower value than the input by 0.36 dex (or 1.5-$\sigma$). In terms of chemical abundances, H$_2$O is off by 0.03 dex ($\sim$1-$\sigma$), CO is consistent with an error bar of 0.10-0.15 dex, CO$_2$ is not detected with a upper limit of -7 dex, and CH$_4$ is consistent with an error bar of 0.10-0.30 dex. This exercise allows us to quantify to what extent we can constrain the mixing ratios and the extinction coefficient $\alpha$. 

Lastly, we allow all parameters to vary. As shown in Table \ref{tab:mock_result}, log(g) is consistent within 1 to 2-$\sigma$ with uncertainty of 0.25 dex. Planet radius is off by 0.6 R$_{\rm{Jupiter}}$ or 10-$\sigma$. The retrieved H$_2$O and CO abundances are consistent with uncertainties between 0.15 and 0.30 dex. The upper limits for CO$_2$ and CH$_4$ are consistent with input values. Cloud species MgSiO$_3$ mixing ratio is off by $\sim$1 dex but with a large error bar of 0.7 dex. Extinction coefficient $\alpha$ has a bias of 0.6 dex.  

\subsection{Lessons Learned from Mock Data}
\label{sec:lessons}

Here we summarize what has been learned from the exercise of retrieving on mock data:
\begin{itemize}
    \item In the absence of modeling systematics, i.e., if using the same model to generate and retrieve the data, we are able to constrain the mixing ratio for molecular species, cloud species MgSiO$_3$, and the extinction coefficient $\alpha$ to describe a featureless spectral slope. However, a bias of $\sim$0.2 dex, 1.0 dex, and 0.6 dex can exist for inferring the values of the mixing ratios for molecular species, MgSiO$_3$, and $\alpha$, respectively. 
    \item In more constraining cases in which molecular species are fixed to chemical equilibrium values and/or some cloud parameters are fixed, typical uncertainties for MgSiO$_3$ and $\alpha$ are 0.2 dex and 0.1-0.3 dex, respectively. 
    \item {{Planet properties such as surface gravity and planet radius are generally retrieved within 0.1 dex and 0.05 R$_{\rm{Jupiter}}$ for constraining cases in which molecular species are fixed. However, retrieved planet properties are unreliable when chemical abundances become free parameters in retrievals. }}
    \item {{By comparing the ``all varying" case of 1x solar metallicity and the case where abundances are fixed at 10x solar metallicity, we can distinguish between 1x and 10x solar metallicity cases at 3-$\sigma$ level using CO and 6-$\sigma$ level using H$_2$O. }}
    \item {{By comparing C/O ratios between fixed and varying cases, the C/O has a typical error bar of 0.1 and a upward bias of $\sim$0.1. }}
\end{itemize}

\begin{figure*}[h]
\begin{tabular}{c}
\includegraphics[width=14.0cm]{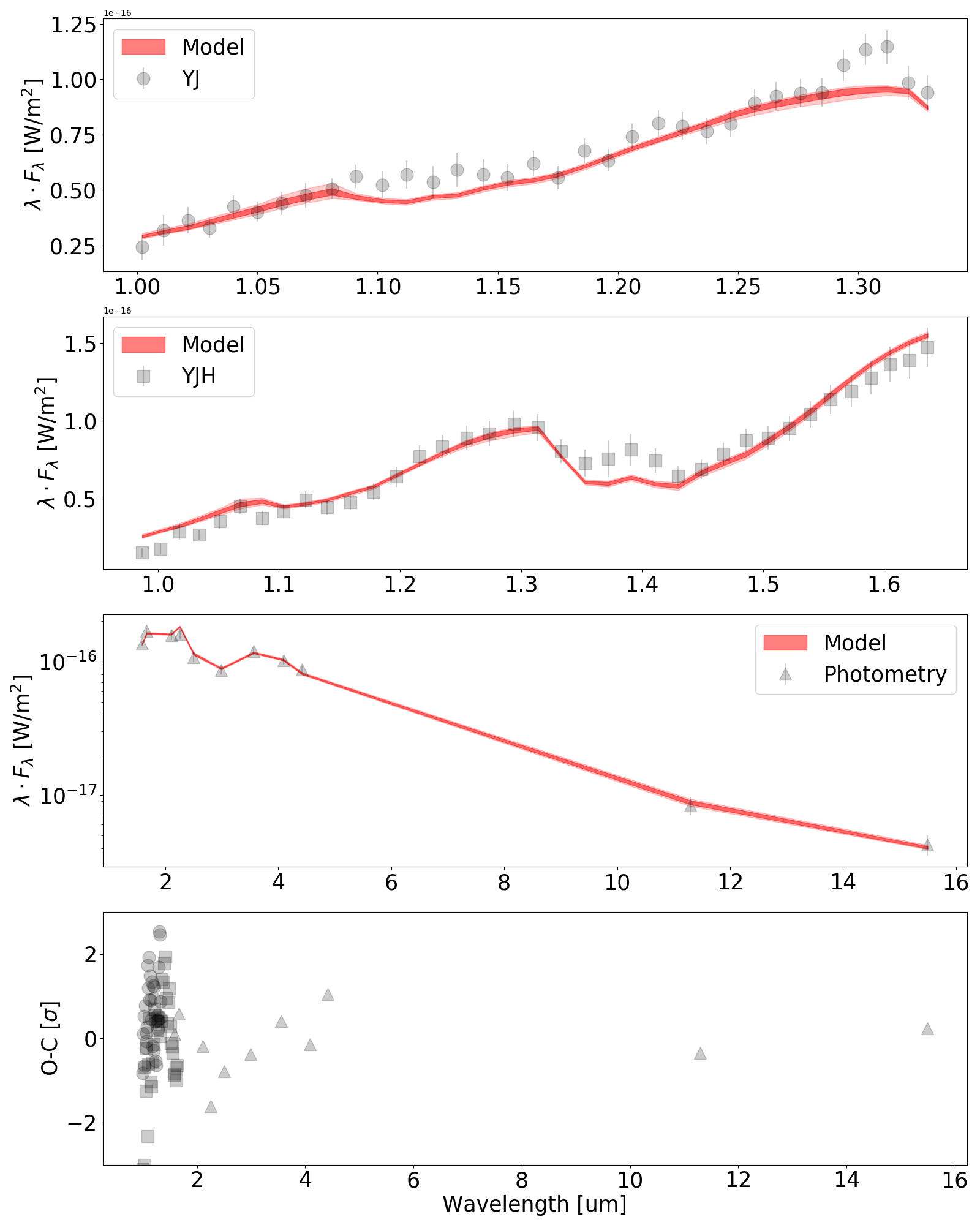}
\end{tabular}
\caption{{\bf{Retrieved spectra for HIP 65426 b (assuming 1x solar metallcity and C/O=0.4)}}. Top three panels are the observed spectroscopic and photometric data (black) and the 1-$\sigma$ (16 to 84 percentile, darker red) and 2-$\sigma$ (2.5 to 97.5 percentile, lighter red) distribution of modeled spectra. The bottom panel is a residual plot with data minus model and divided by the individual errors. {{More results with other assumptions can be found in Table \ref{tab:real_result}}}.  
\label{fig:data_model}}
\end{figure*} 

\begin{figure*}[ht]
\epsscale{1.1}
\plotone{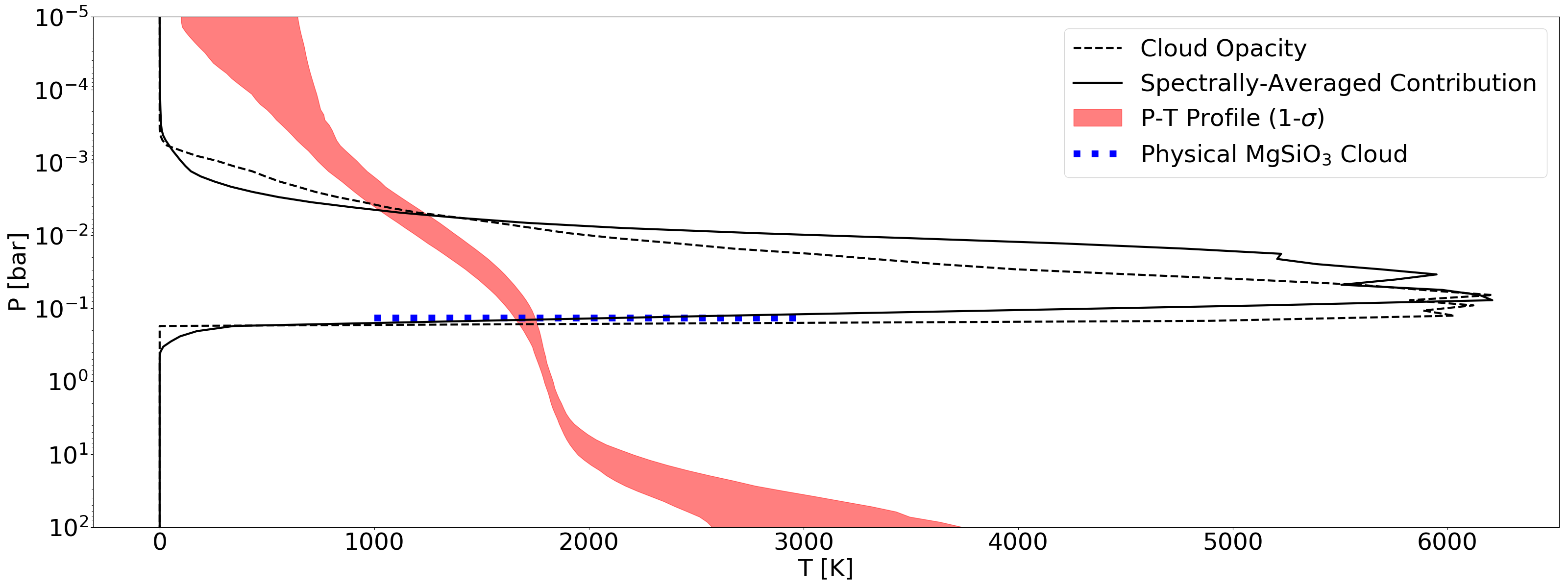}
\caption{{\bf{Retrieved P-T profile (1-$\sigma$ region in red shaded region) for HIP 65426 b (assuming 1x solar metallcity and C/O=0.4)}} The spectrally-averaged contribution function is shown as the black solid line, which completely overlaps with retrieved cloud layer (black dashed line). The pressure level of the retrieved cloud layer is consistent with that of a physical MgSiO$_3$ cloud (blue dotted line). {{More results with other assumptions can be found in Table \ref{tab:real_result}}}.
\label{fig:pt}}
\end{figure*} 

\section{Retrievals on SPHERE+JWST Joint Data}
\label{sec:results}

\subsection{{{Solar}} Metallicity and C/O$<$1 for HIP 65426~\lowercase{b}}
\label{sec:solar_co}
We perform retrieval analyses for the joint data set of SPHERE and JWST data with different combination of C/O ratios (0.40, 0.55, and 1.00) and metallicities (1x and 10x solar). We also run retrieval by varying all parameters. The results are given in Table \ref{tab:real_result}. 

By comparing Bayesian evidence (EV), the retrieval model using sub-solar C/O (C/O = 0.40) and solar metallicity has the highest Bayesian evidence. However, the next most preferred model using solar C/O (C/O = 0.55) and solar metallicity has a differential natural log evidence ($\Delta\ln(EV)$ of -1.47. This suggests that the both these two models are consistent with data~\citep{Benneke2013,Trotta2008}. {{Therefore, the data can be explained by a model with 1x solar metallicity and two C/O ratios at 0.40 and 0.55.  }}

Other models are strongly disfavored. For example, {{models with}} 10x solar metallicity at all C/O ratios have $\Delta\ln(EV)$ ranging from {{-15.86 to -365.29}}. In addition, models with {{a high C/O ratio (C/O = 1.0)}} are disfavored as well as the retrieval results by varying all parameters. Our results are broadly consistent with findings in~\citet{Petrus2021}. More details on the comparison to~\citet{Petrus2021} are given in \S \ref{sec:petrus}.

\subsection{Strong Evidence of Silicate Clouds}
\label{sec:silicate}
We find strong evidence for the presence of silicate clouds. {{First, all retrievals return a mixing ratio for MgSiO$_3$ that is at least -3.4, a value that we can confidently detect for mock data (\S \ref{sec:test}).}} The {{detection}} is robust against all assumptions we consider. Second, the dip in the 11.4 $\mu$m photometry (Fig. \ref{fig:data_model}) is another visual evidence for the silicate clouds. The inferred silicate clouds also play an important role in regulating the emerging flux. This is because the peak pressure levels ($\sim$0.1 bar to 5 mbar) of the spectrally-averaged contribution function overlaps with {{that}} of the cloud opacity distribution (Fig. \ref{fig:pt}). 

\subsection{No Evidence of Featureless {{Dust}} Extinction}
\label{sec:ext_coe}
While HIP 65426 b has an older age estimate~\citep[$14\pm4$ Myr][]{Chauvin2017} when compared to that of PDS 70 b and c~\citep[5-8 Myr][]{Wang2021b,Keppler2018}, we would like to investigate the possibility of HIP 65426 b processing an enshrouding dust that produces a featureless extinction spectral slope, similar to that as inferred from the PDS 70 planets~\citep{Wang2021b}. 

We do not find any evidence of such an enshrouding dust. All but two cases with C/O = 1.00 have $\alpha$ consistent with zero {{within 2-$\sigma$}} (Table \ref{tab:real_result}). Although the two cases with C/O = 1.00 infer a non-zero $\alpha$, they have the lowest evidence and therefore can be safely discarded. It is not surprising to find a zero $\alpha$ because of the older age of HIP 65426 b than PDS 70 b and c. HIP 65426 b may have already ceased the accretion and the dust has already settled a few Myrs after the active accretion. 

\subsection{Non-Sensible Chemical Abundances}
\label{sec:non_sense}

\begin{figure}[h]
\includegraphics[width=8.5cm]{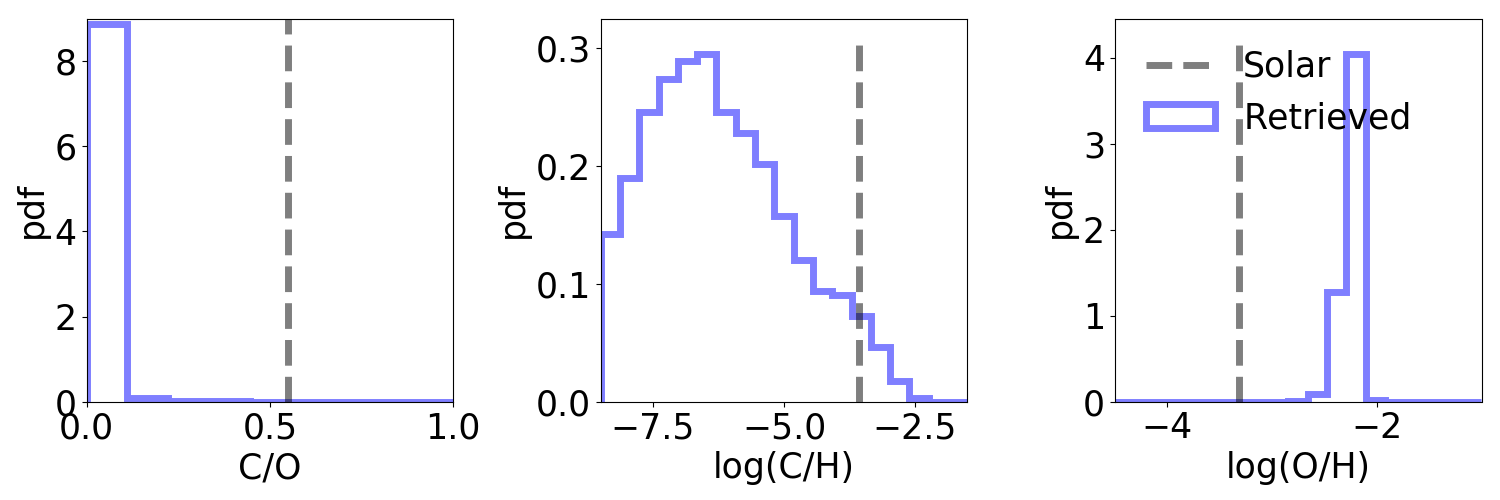} 
\caption{{\bf{Retrieved C and O abundances and C/O}}: C and O abundances from posterior samples are blue histograms, and stellar values are the black dashed lines. Free retrieval by varying all parameters returns non-physical results {{(e.g., C/O$\sim$0)}}. More details can be found in \S \ref{sec:non_sense}. 
\label{fig:co_comp}}  
\end{figure}

We find that the retrieved mixing ratios for molecules (Table \ref{tab:real_result} ``All Varying" case) are significantly different from the chemical equilibrium cases that we have considered. For example, mixing ratio for H$_2$O is at -1.1 dex, which is {{$>$3-$\sigma$}} off the fixed mixing ratios. In addition, the retrieved H$_2$O mixing ratio is unreasonably high and at the edge of prior range. Moreover, {{CO mixing ratio is too low}} compared to the fixed equilibrium values (Table \ref{tab:real_result}) {{and the retrieved C/O for the ``All Varying" case peaks at 0.0 (Fig. \ref{fig:co_comp}). }}

Fig. \ref{fig:co_comp} shows the retrieved C and O abundances and C/O for the all-varying retrieval run. The retrieved C/O is at 0 indicating an abnormally high O and low C abundance. This is driven by the aforementioned high H$_2$O mixing ratio {{and low CO mixing ratio}}. In addition, the Bayesian evidence for the all-varying run has a $\Delta\ln(EV)$ of {{-20.43}} as compared to the preferred model. Therefore, the all-varying run and the inferred non-physical molecular mixing ratios should be regarded with extremely low confidence.

\section{Discussion}
\label{sec:discussion}

\subsection{Comparison to~\citet{Petrus2021}}
\label{sec:petrus}

\citet{Petrus2021} conducted a thorough analysis of the joint data set of VLT/SINFONI, SHERE, and NaCo that covers $\sim$1-5 $\mu$m. Here we compare our result of the run with the highest evidence to their ``K band with continuum" run in their Table 2 because the run is a similar retrieval analysis and has the most inferred parameters to compare with. 

Planet bulk properties such as log(g) and R$_P$ are consistent. Our inferred log(g) and R$_P$ are {{$3.19^{+0.23}_{-0.25}$}} and {{$1.23^{+0.02}_{-0.02}$}} vs. $<$4.2 and $1.28^{+0.10}_{-0.11}$ in their Table 2. {{Note that}} error bars and systematics for log(g) and R$_P$ in our retrieval analysis can be as high as {{0.6}} dex and {{0.1}} R$_{\rm{Jupiter}}$ for the fixed abundance cases as learned from \S \ref{sec:test}, which make our results consistent with those in~\citet{Petrus2021}. 

{{In terms of chemical properties, we find that models with solar metallicity, solar C/O (C/O = 0.55) or sub-solar C/O (C/O = 0.40) are the most likely models ($\Delta\ln(EV)$ = -1.47). }}This is consistent with the finding in~\citet{Petrus2021} that the planet metallicity {{[M/H]}} is $0.05^{+0.24}_{-0.22}$ and C/O is lower than 0.50, especially given that our models with C/O = 1.00 have the lowest evidence ($\Delta\ln(EV)<-300$, Table \ref{tab:real_result}). 

{{Our retrieved effective temperature T$_{\rm{eff}}$ is $1477^{+8}_{-10}$ K.}} In comparison, T$_{\rm{eff}}$ in~\citet{Petrus2021} is $1518^{+88}_{-71}$ K, which is consistent within 1-$\sigma$. To obtain T$_{\rm{eff}}$ and the associated uncertainties, we integrate flux from 0.9 $\mu$m to 17.0 $\mu$m using modeled spectra with randomly drawn posteriors to get luminosity. Then we use Stefan-Boltzmann law and the planet radius posterior samples to calculate T$_{\rm{eff}}$. {{Our inferred planet luminosity is $\log(L/L_\odot) = -4.167^{+0.003}_{-0.002}$}}, which is also consistent with the result in~\citet{Petrus2021} at $\log(L/L_\odot) = -4.10\pm0.2$. 


\subsection{Comparison to~\citet{Carter2022}}
\label{sec:carter}

~\citet{Carter2022} inferred planet properties using two methods: fitting evolutionary tracks and atmospheric models. The two methods return significantly different results. Fitting evolutionary tracks results in a planet radius of $1.45\pm0.03$ R$_{\rm{Jupiter}}$, log(g) of $3.93\pm0.07$, and T$_{\rm{eff}}$ of $1282^{+26}_{-31}$ K. Fitting atmospheric models results in a planet radius of $0.92\pm0.04$ R$_{\rm{Jupiter}}$, log(g) of $4.07\pm0.19$, and T$_{\rm{eff}}$ of $1667^{+25}_{-24}$ K. They concluded that the result by fitting  evolutionary tracks is more physical because a {{larger}} planet radius is expected for a young contracting planet. Our result in this work is different from either of their results. This can be attributed to the different approach: we conduct retrieval analysis whereas~\citet{Carter2022} fit the same data set with evolutionary tracks and atmospheric models. 

\subsection{Potential Future Improvement}
\label{sec:future}

There are a few drawbacks of performing retrieval analyses and below we lay out certain aspects that can be improved in the future. While we find strong evidence for silicate clouds, current retrieval analysis does not allow us to make more physical and quantitative statement about the silicate clouds. As pointed out in~\citet{Molliere2020}, using $\log$(mr$_{\rm{MgSiO}_3}$), $\log$(K$_{zz}$), $f_{sed}$, and $\sigma_g$ are just ``a glorified way" of parameterizing clouds. These parameters, however, are not self-consistently included in the model. For example, cloud feedback~\citep{Tan2021, Tan2021b} is not considered in \petit. Self-inconsistency like this prevents us from quantitatively and accurately interpret the physical cloud properties. {{In addition, while MgSiO$_3$ is the only cloud species that is considered this work, there can be other unaccounted cloud species~\citep[e.g., {{SiO$_2$, Fe, and Mg$_2$SiO$_4$}}, ][]{Burningham2021} to condensate around $\sim$1500 K, which is the effective temperature of HIP 65426 b. }}

\S \ref{sec:non_sense} already discusses the unrealistic chemistry that is inferred from the all-varying run. This can be largely attributed to the lack of data with higher spectral resolution: the JWST and SPHERE data that are used in this work are either photometric data or IFU data with R lower than $\sim$15. Including data with higher spectral resolution such as VLT/SINFONI (R$\sim$5,500), JWST (e.g., NIRSpec with R up  to 3,600), and {{VLT/HiRISE~\citep[R$\sim$100,000, ][]{Vigan2018,Morsy2022}}} will resolve molecular lines and offer a much better direct measurement and constraint on chemical abundances.  

\section{Summary}
\label{sec:summary}
We perform retrieval analyses on a joint data set of SPHERE and JWST for HIP 65426 b. {{We find that the atmosphere of HIP 65426 b is more likely to have a solar metallicity and a C/O ratio at 0.4 or 0.55 than 10x solar metallicity and a C/O=1.0 }}based on model comparison using Bayesian evidence. The preferred model shows strong evidence of silicate clouds and the presence of silicate clouds is robust against all assumptions that we consider in this work. We find no sign of an enshrouding dust for HIP 65426 b that exists in other young planets such as PDS 70 b and c. Below we summarize {{our findings from retrieval analyses}} with low-resolution IFU data and JWST NIRCam and MIRI photometric data points. 
\begin{itemize}
    \item Low-resolution and photometric data points that cover a broad wavelength range can provide a certain level of constraint on metallicity and C/O ratio, {{e.g. $>$3-$\sigma$ to distinguish between 1 dex difference in metallicity (i.e., 1x and 10x solar metallicity) and a few tenths in C/O (\S \ref{sec:lessons}). The case study on HIP 65426 b data set suggests}} that retrieval {{analyses be}} done with the guidance of an equilibrium chemistry model with a simple quenching mechanism (\S \ref{sec:solar_co}). However, a free retrieval that varies mixing ratios for molecules can usually lead to unrealistic chemistry (\S \ref{sec:non_sense} and Fig. \ref{fig:co_comp}).  
    \item The presence of clouds and the type of cloud can be inferred by comparing the pressure range for the cloud opacity and the pressure range for the flux contribution function (Fig. \ref{fig:pt}), and by checking the dip of the 11.4 $\mu$m photometric data point (Fig. \ref{fig:data_model} and \S \ref{sec:silicate}). These features usually correspond to a high inferred MgSiO$_3$ mixing ratio {{(Table \ref{tab:real_result})}}, which again points to the presence of silicate clouds. However, quantitatively and accurately interpreting the cloud physics and chemistry will require self-consistent models which are not currently included in any retrieval codes (\S \ref{sec:future}).  
    \item {{Our work suggests using}} mock data to test the retrieval code and understand the limitations of the actual data set, the mock data set, and the retrieval analysis (\S \ref{sec:test}). Blind and brutal application usually results in an underestimation of error bars, biases, and systematics in the retrieval analysis (\S \ref{sec:lessons} and Table \ref{tab:mock_result}). 
\end{itemize}

\noindent
{\bf{Acknowledgments}} We thank the anonymous referee for the comments and suggestions to significantly improve the paper. This work is supported by the National Science Foundation under Grant No. 2143400. 

\bibliography{sample63}{}
\bibliographystyle{aasjournal}



\begin{deluxetable}{lcccc}
\tabletypesize{\tiny}
\tablewidth{0pt}
\tablecaption{Parameters used in retrieval and their priors.\label{tab:prior}}
\tablehead{
\colhead{\textbf{Parameter}} &
\colhead{\textbf{Unit}} &
\colhead{\textbf{Type}} &
\colhead{\textbf{Lower}} &
\colhead{\textbf{Upper}} \\
\colhead{\textbf{}} &
\colhead{\textbf{}} &
\colhead{\textbf{}} &
\colhead{\textbf{or Mean}} &
\colhead{\textbf{or Std}} 
}

\startdata
Surface Gravity ($\log$(g))                   &  cgs                 &   Uniform        &  2.5        &   5.5       \\
Planet Radius (R$_P$)                       &  M$_{\rm{Jupiter}}$  &   Uniform        &   0.5      &   5.0       \\
H$_2$O Mixing Ratio ($\log$(mr$_{\rm{H}_2\rm{O}}$)) &  \nodata             &   Log-uniform      &  -10       &   -1      \\
CO Mixing Ratio ($\log$(mr$_{\rm{C}\rm{O}}$))       &  \nodata             &   Log-uniform       &   -10     &   -1      \\
CO$_2$ Mixing Ratio ($\log$(mr$_{\rm{C}\rm{O}_2}$)) &  \nodata             &   Log-uniform       &  -10       &   -1      \\
CH$_4$ Mixing Ratio ($\log$(mr$_{\rm{C}\rm{H}_4}$)) &  \nodata             &   Log-uniform       &  -10       &   -1      \\
Temperature at 3.2 bar (t$_{\rm{int}}$)      &  K                   &   Uniform   &   800 &   2500  \\
$\Delta T$ between 100 and 32 bar      &  K                   &   Uniform   &   0 &   2500  \\
$\Delta T$ between 32 and 10 bar      &  K                   &   Uniform   &   0 &   2000  \\
$\Delta T$ between 10 and 3.2 bar      &  K                   &   Uniform   &   0 &   1500  \\
$\Delta T$ between 3.2 and 1 bar      &  K                   &   Uniform   &   0 &   1000  \\
$\Delta T$ between 1 and 0.1 bar      &  K                   &   Uniform   &   0 &   1000  \\
$\Delta T$ between 0.1 bar and 1 mbar      &  K                   &   Uniform   &   0 &   1000  \\
$\Delta T$ between 1 mbar and 10 nbar      &  K                   &   Uniform   &   0 &   1000  \\
MgSiO$_3$ Mixing Ratio ($\log$(mr$_{\rm{MgSiO}_3}$)) &  \nodata             &   Log-uniform      &  -10       &   -1      \\
Vertical diffusion coefficient ($\log$(K$_{zz}$))           &  cm$2\cdot$s$^{-1}$                 &   Log-uniform       &  5        &   10       \\
$v_{\rm{settling}}/v_{\rm{mixing}}$ ($f_{sed}$)          &  \nodata                 &   Uniform       &  0        &   5       \\
Width of log-normal particle size distribution ($\sigma_g$))           &  \nodata                 &   Uniform       &  1.05        &   3.05       \\
Extinction coefficient ($\alpha$)           &  \nodata                 &   Uniform       &  0.0        &   5.0       \\
\enddata



\end{deluxetable}

\clearpage
\newpage




\begin{deluxetable}{lcccccccc}
\rotate

\tabletypesize{\scriptsize}
\tablewidth{0pt}
\tablecaption{Input and Retrieved Parameters Using Mock Data.\label{tab:mock_result}}
\tablehead{
\colhead{\textbf{Parameter}} &
\colhead{\textbf{Unit}} &
\multicolumn{2}{c}{\textbf{1x solar}} & 
\multicolumn{2}{c}{\textbf{10x solar}} &
& \colhead{\textbf{All Varying}} & \colhead{\textbf{Input}} \\
\multicolumn{2}{c}{\textbf{MgSiO$_3$}} &
\colhead{\textbf{low$^{\ast}$}} &
\colhead{\textbf{high}} &
\colhead{\textbf{low}} &
\colhead{\textbf{high}} & \colhead{\textbf{varying}} & &
}


\startdata
$\log$(g)                                & cgs                  & $      4.03^{     +0.03}_{     -0.02}$ &     $      3.99^{     +0.12}_{     -0.06}$ &     $      4.14^{     +0.11}_{     -0.12}$ &     $      3.39^{     +0.16}_{     -0.12}$ &     $      3.17^{     +0.08}_{     -0.13}$ &     $      4.33^{     +0.24}_{     -0.22}$   & 4.0    \\
R$_P$                                    & R$_{\rm{Jupiter}}$   & $      3.53^{     +0.01}_{     -0.01}$ &     $      3.46^{     +0.03}_{     -0.02}$ &     $      3.54^{     +0.02}_{     -0.02}$ &     $      3.38^{     +0.03}_{     -0.03}$ &     $     3.50$ &     $      2.89^{     +0.06}_{     -0.06}$    & 3.5  \\
$\log$(mr$_{\rm{H}_2\rm{O}}$)            & \nodata              & \multicolumn{2}{c}{-2.605} &     \multicolumn{2}{c}{-1.655} &     $    -2.635^{    +0.031}_{    -0.040}$ &     $    -2.549^{    +0.142}_{    -0.143}$   & -2.605 or -1.655   \\
$\log$(mr$_{\rm{C}\rm{O}}$)              & \nodata              & \multicolumn{2}{c}{-2.258} &     \multicolumn{2}{c}{-1.303} &     $    -2.169^{    +0.126}_{    -0.111}$ &     $    -2.051^{    +0.331}_{    -0.300}$   & -2.258 or -1.303   \\
$\log$(mr$_{\rm{C}\rm{O}_2}$)            & \nodata              & \multicolumn{2}{c}{-6.300} &     \multicolumn{2}{c}{-4.337} &     $    -8.575^{    +0.910}_{    -0.837}$ &     $    -8.420^{    +1.019}_{    -0.950}$   & -6.300 or -4.337   \\
$\log$(mr$_{\rm{C}\rm{H}_4}$)            & \nodata              & \multicolumn{2}{c}{-5.669} &     \multicolumn{2}{c}{-5.661} &     $    -5.905^{    +0.105}_{    -0.280}$ &     $    -7.342^{    +1.205}_{    -1.598}$   & -5.669 or -5.661   \\
t$_{\rm{int}}$                           & K                    & $      1795^{       +54}_{       -70}$ &     $      1771^{       +75}_{       -77}$ &     $      1838^{      +102}_{      -128}$ &     $      2349^{       +95}_{      -137}$ &     $      1677^{       +69}_{       -70}$ &     $      1618^{       +72}_{      -102}$      & 1800 \\
$\Delta T$ between 100 and 32 bar        & K                    & $      1512^{      +639}_{      -630}$ &     $      1860^{      +471}_{      -926}$ &     $      1374^{      +769}_{      -891}$ &     $      1254^{      +831}_{      -845}$ &     $      1765^{      +462}_{      -669}$ &     $      2107^{      +274}_{      -459}$      & 2000 \\
$\Delta T$ between 32 and 10 bar         & K                    & $      1639^{      +249}_{      -290}$ &     $      1776^{      +161}_{      -282}$ &     $      1237^{      +501}_{      -646}$ &     $      1089^{      +613}_{      -717}$ &     $      1691^{      +190}_{      -269}$ &     $      1501^{      +310}_{      -372}$      & 1500 \\
$\Delta T$ between 10 and 3.2 bar        & K                    & $       954^{      +196}_{      -153}$ &     $       985^{      +305}_{      -343}$ &     $       541^{      +332}_{      -241}$ &     $      1082^{      +296}_{      -469}$ &     $       631^{      +260}_{      -225}$ &     $       176^{      +257}_{      -124}$      & 1000 \\
$\Delta T$ between 3.2 and 1 bar         & K                    & $       517^{       +86}_{      -108}$ &     $       331^{      +118}_{      -111}$ &     $       715^{      +117}_{      -156}$ &     $       845^{       +95}_{      -132}$ &     $       330^{      +101}_{      -104}$ &     $       168^{      +167}_{      -107}$      & 500 \\
$\Delta T$ between 1 and 0.1 bar         & K                    & $       479^{       +72}_{       -59}$ &     $       687^{       +75}_{       -77}$ &     $       311^{       +66}_{       -50}$ &     $       553^{       +45}_{       -44}$ &     $       475^{       +73}_{       -66}$ &     $       496^{      +196}_{      -195}$      & 500 \\
$\Delta T$ between 0.1 bar and 1 mbar    & K                    & $       547^{      +260}_{      -250}$ &     $       360^{      +272}_{      -214}$ &     $       766^{      +150}_{      -187}$ &     $       618^{       +62}_{       -60}$ &     $       543^{      +257}_{      -282}$ &     $       540^{      +288}_{      -331}$      & 500 \\
$\Delta T$ between 1 mbar and 10 nbar    & K                    & $       511^{      +332}_{      -338}$ &     $       515^{      +324}_{      -341}$ &     $       473^{      +351}_{      -319}$ &     $       543^{      +308}_{      -348}$ &     $       514^{      +285}_{      -302}$ &     $       512^{      +296}_{      -304}$      & 500 \\
$\log$(mr$_{\rm{MgSiO}_3}$)              & \nodata              & $     -4$ &     $     -3$ &     $     -4$ &     $     -3$ &     $     -3.36^{     +0.11}_{     -0.23}$ &     $     -4.07^{     +0.68}_{     -2.90}$   &   -4 or -3 \\
$\log$(K$_{zz}$)                         & cm$2\cdot$s$^{-1}$   & \multicolumn{5}{c}{8.0} &     $      8.04^{     +0.98}_{     -1.10}$      & 8.0 \\
$f_{sed}$                                & \nodata              & \multicolumn{5}{c}{1.3} &     $      3.36^{     +0.97}_{     -1.34}$      & 1.3 \\
Cloud log-normal size $\sigma_g$         & \nodata              & \multicolumn{5}{c}{1.31} &     $      2.06^{     +0.56}_{     -0.56}$      & 1.31 \\
Extinction coefficient $\alpha$          & \nodata              & $      3.10^{     +0.05}_{     -0.05}$ &     $      3.34^{     +0.16}_{     -0.16}$ &     $      2.99^{     +0.22}_{     -0.23}$ &     $      2.53^{     +0.32}_{     -0.26}$ &     $      2.96^{     +0.23}_{     -0.14}$ &     $      3.62^{     +0.29}_{     -0.25}$      & 3.0 \\
C/O &\nodata&0.55&0.55&0.55&0.55&$      0.65^{     +0.05}_{     -0.05}$&$      0.66^{     +0.09}_{     -0.10}$& 0.55\\
\enddata

\tablecomments{{$\ast$: Low, high, and varying refer to cases with low (-4), high (-3), and varying mixing ratio for MgSiO$_3$; see \S \ref{sec:analyze_mock} for more details of each case (or column). }}

\end{deluxetable}

\clearpage
\newpage

\startlongtable
\begin{deluxetable}{cccc}
\tabletypesize{\tiny}
\tablewidth{0pt}
\tablecaption{Joint Data for JWST and SPHERE \label{tab:data}}
\tablehead{
\colhead{\textbf{Filter$^\ast$}} &
\colhead{\textbf{$\lambda_0$}} &
\colhead{\textbf{$\Delta\lambda$}} &
\colhead{\textbf{flux}} \\
\colhead{\textbf{}} &
\colhead{\textbf{[$\mu$m]}} &
\colhead{\textbf{[$\mu$m]}} &
\colhead{\textbf{[W/m/$\mu$m]}}
}

\startdata
H2             &1.588          &0.053          &$8.57           \pm0.38           \times10^{-17}$\\
H3             &1.667          &0.055          &$10.13          \pm0.56           \times10^{-17}$\\
K1             &2.102          &0.102          &$7.50           \pm0.60           \times10^{-17}$\\
K2             &2.255          &0.109          &$7.10           \pm0.60           \times10^{-17}$\\
F250M          &2.500          &0.180          &$4.29           \pm0.33           \times10^{-17}$\\
F300M          &2.990          &0.310          &$2.89           \pm0.20           \times10^{-17}$\\
F356M          &3.560          &0.780          &$3.36           \pm0.23           \times10^{-17}$\\
F410M          &4.090          &0.430          &$2.49           \pm0.18           \times10^{-17}$\\
F444M          &4.420          &1.020          &$1.97           \pm0.13           \times10^{-17}$\\
F1140C         &11.300         &1.600          &$7.40           \pm1.16           \times10^{-19}$\\
F1550C         &15.500         &1.800          &$2.74           \pm0.46           \times10^{-19}$\\
               &1.002          &0.011          &$2.43           \pm0.57           \times10^{-17}$\\
               &1.011          &0.011          &$3.15           \pm0.67           \times10^{-17}$\\
               &1.021          &0.011          &$3.56           \pm0.57           \times10^{-17}$\\
               &1.030          &0.011          &$3.20           \pm0.43           \times10^{-17}$\\
               &1.040          &0.011          &$4.09           \pm0.48           \times10^{-17}$\\
               &1.050          &0.011          &$3.83           \pm0.43           \times10^{-17}$\\
               &1.060          &0.011          &$4.15           \pm0.49           \times10^{-17}$\\
               &1.070          &0.011          &$4.46           \pm0.52           \times10^{-17}$\\
               &1.081          &0.011          &$4.68           \pm0.43           \times10^{-17}$\\
               &1.091          &0.011          &$5.14           \pm0.48           \times10^{-17}$\\
               &1.102          &0.011          &$4.75           \pm0.54           \times10^{-17}$\\
               &1.112          &0.011          &$5.13           \pm0.58           \times10^{-17}$\\
               &1.123          &0.011          &$4.78           \pm0.64           \times10^{-17}$\\
               &1.133          &0.011          &$5.23           \pm0.68           \times10^{-17}$\\
               &1.144          &0.011          &$4.98           \pm0.61           \times10^{-17}$\\
               &1.154          &0.011          &$4.82           \pm0.52           \times10^{-17}$\\
               &1.165          &0.011          &$5.32           \pm0.49           \times10^{-17}$\\
               &1.175          &0.011          &$4.74           \pm0.45           \times10^{-17}$\\
               &1.186          &0.011          &$5.71           \pm0.48           \times10^{-17}$\\
               &1.196          &0.011          &$5.29           \pm0.41           \times10^{-17}$\\
               &1.206          &0.011          &$6.16           \pm0.47           \times10^{-17}$\\
               &1.217          &0.011          &$6.59           \pm0.49           \times10^{-17}$\\
               &1.227          &0.011          &$6.43           \pm0.50           \times10^{-17}$\\
               &1.237          &0.011          &$6.20           \pm0.48           \times10^{-17}$\\
               &1.247          &0.011          &$6.41           \pm0.48           \times10^{-17}$\\
               &1.257          &0.011          &$7.10           \pm0.48           \times10^{-17}$\\
               &1.266          &0.011          &$7.29           \pm0.50           \times10^{-17}$\\
               &1.276          &0.011          &$7.34           \pm0.50           \times10^{-17}$\\
               &1.285          &0.011          &$7.32           \pm0.50           \times10^{-17}$\\
               &1.294          &0.011          &$8.21           \pm0.54           \times10^{-17}$\\
               &1.303          &0.011          &$8.70           \pm0.54           \times10^{-17}$\\
               &1.312          &0.011          &$8.73           \pm0.58           \times10^{-17}$\\
               &1.321          &0.011          &$7.45           \pm0.59           \times10^{-17}$\\
               &1.329          &0.011          &$7.07           \pm0.58           \times10^{-17}$\\
               &0.987          &0.019          &$1.56           \pm0.31           \times10^{-17}$\\
               &1.002          &0.019          &$1.79           \pm0.35           \times10^{-17}$\\
               &1.018          &0.019          &$2.85           \pm0.51           \times10^{-17}$\\
               &1.034          &0.019          &$2.62           \pm0.30           \times10^{-17}$\\
               &1.051          &0.019          &$3.39           \pm0.44           \times10^{-17}$\\
               &1.068          &0.019          &$4.24           \pm0.46           \times10^{-17}$\\
               &1.086          &0.019          &$3.47           \pm0.41           \times10^{-17}$\\
               &1.104          &0.019          &$3.80           \pm0.41           \times10^{-17}$\\
               &1.122          &0.019          &$4.40           \pm0.50           \times10^{-17}$\\
               &1.140          &0.019          &$3.91           \pm0.40           \times10^{-17}$\\
               &1.159          &0.019          &$4.13           \pm0.42           \times10^{-17}$\\
               &1.178          &0.019          &$4.64           \pm0.45           \times10^{-17}$\\
               &1.197          &0.019          &$5.37           \pm0.54           \times10^{-17}$\\
               &1.216          &0.019          &$6.36           \pm0.57           \times10^{-17}$\\
               &1.235          &0.019          &$6.76           \pm0.60           \times10^{-17}$\\
               &1.255          &0.019          &$7.11           \pm0.62           \times10^{-17}$\\
               &1.274          &0.019          &$7.23           \pm0.63           \times10^{-17}$\\
               &1.294          &0.019          &$7.60           \pm0.67           \times10^{-17}$\\
               &1.314          &0.019          &$7.30           \pm0.65           \times10^{-17}$\\
               &1.333          &0.019          &$6.05           \pm0.58           \times10^{-17}$\\
               &1.353          &0.019          &$5.40           \pm0.65           \times10^{-17}$\\
               &1.372          &0.019          &$5.52           \pm0.87           \times10^{-17}$\\
               &1.391          &0.019          &$5.88           \pm0.73           \times10^{-17}$\\
               &1.411          &0.019          &$5.29           \pm0.55           \times10^{-17}$\\
               &1.430          &0.019          &$4.52           \pm0.46           \times10^{-17}$\\
               &1.449          &0.019          &$4.77           \pm0.44           \times10^{-17}$\\
               &1.467          &0.019          &$5.37           \pm0.48           \times10^{-17}$\\
               &1.486          &0.019          &$5.89           \pm0.51           \times10^{-17}$\\
               &1.504          &0.019          &$5.93           \pm0.50           \times10^{-17}$\\
               &1.522          &0.019          &$6.26           \pm0.52           \times10^{-17}$\\
               &1.539          &0.019          &$6.78           \pm0.56           \times10^{-17}$\\
               &1.556          &0.019          &$7.32           \pm0.60           \times10^{-17}$\\
               &1.573          &0.019          &$7.56           \pm0.62           \times10^{-17}$\\
               &1.589          &0.019          &$8.04           \pm0.67           \times10^{-17}$\\
               &1.605          &0.019          &$8.50           \pm0.71           \times10^{-17}$\\
               &1.621          &0.019          &$8.58           \pm0.71           \times10^{-17}$\\
               &1.636          &0.019          &$9.01           \pm0.76           \times10^{-17}$\\
               &0.000          &0.000          &$0.00           \pm0.00           \times10^{-17}$\\
\enddata

\tablecomments{{$\ast$: {{$H$ and $K$ photometric data are from~\citet{Cheetham2019}; JWST photometric data (beginning with ``F") are from~\citet{Carter2022}; rows without a filter name are data from SPHERE integrated field unit~\citep{Chauvin2017}.}}   }}

\end{deluxetable}

\clearpage
\newpage

\begin{deluxetable}{lcccccccc}
\rotate

\tabletypesize{\scriptsize}
\tablewidth{0pt}
\tablecaption{Retrieved Parameters Using Real Data.\label{tab:real_result}}
\tablehead{
\colhead{\textbf{Parameter}} &
\colhead{\textbf{Unit}} &
\multicolumn{2}{c}{\textbf{C/O = 0.40}} & 
\multicolumn{2}{c}{\textbf{C/O = 0.55}} & 
\multicolumn{2}{c}{\textbf{C/O = 1.00}} &  \\
\multicolumn{2}{l}{\textbf{Solar Metallicity}} &
\colhead{\textbf{1x}} &
\colhead{\textbf{10x}} &
\colhead{\textbf{1x}} &
\colhead{\textbf{10x}} & 
\colhead{\textbf{1x}} &
\colhead{\textbf{10x}} & 
\colhead{\textbf{All Varying}}
}

\startdata
$\log$(g)                                & cgs                  & $      3.19^{     +0.23}_{     -0.25}$ &     $      3.41^{     +0.19}_{     -0.25}$ &     $      3.03^{     +0.21}_{     -0.21}$ &     $      3.57^{     +0.15}_{     -0.15}$ &     $      2.63^{     +0.12}_{     -0.08}$ &     $      2.60^{     +0.09}_{     -0.06}$ &     $      2.60^{     +0.09}_{     -0.06}$      \\
R$_P$                                    & R$_{\rm{Jupiter}}$   & $      1.23^{     +0.02}_{     -0.02}$ &     $      1.23^{     +0.02}_{     -0.02}$ &     $      1.25^{     +0.02}_{     -0.02}$ &     $      1.25^{     +0.02}_{     -0.02}$ &     $      1.18^{     +0.03}_{     -0.02}$ &     $      1.14^{     +0.02}_{     -0.02}$ &     $      1.18^{     +0.01}_{     -0.01}$      \\
$\log$(mr$_{\rm{H}_2\rm{O}}$)            & \nodata              & $     -2.311$ &     $     -1.366$ &     $     -2.605$ &     $     -1.655$ &     $     -4.314$ &     $     -3.994$ &     $     -1.10^{     +0.06}_{     -0.09}$      \\
$\log$(mr$_{\rm{C}\rm{O}}$)              & \nodata              & $     -2.259$ &     $     -1.312$ &     $     -2.258$ &     $     -1.303$ &     $     -2.290$ &     $     -1.322$ &     $     -5.91^{     +2.45}_{     -2.30}$      \\
$\log$(mr$_{\rm{C}\rm{O}_2}$)            & \nodata              & $     -5.988$ &     $     -4.036$ &     $     -6.300$ &     $     -4.337$ &     $     -8.954$ &     $     -7.475$ &     $     -7.23^{     +1.41}_{     -1.45}$      \\
$\log$(mr$_{\rm{C}\rm{H}_4}$)            & \nodata              & $     -5.913$ &     $     -5.957$ &     $     -5.669$ &     $     -5.661$ &     $     -3.913$ &     $     -3.331$ &     $     -7.10^{     +1.73}_{     -1.58}$      \\
t$_{\rm{int}}$                           & K                    & $      1853^{       +38}_{       -38}$ &     $      1906^{      +115}_{       -71}$ &     $      1842^{       +34}_{       -36}$ &     $      1849^{       +70}_{       -41}$ &     $      2487^{        +8}_{       -21}$ &     $      2485^{        +9}_{       -21}$ &     $      2244^{      +141}_{      -165}$      \\
$\Delta T$ between 100 and 32 bar        & K                    & $      1206^{      +742}_{      -720}$ &     $      1212^{      +633}_{      -614}$ &     $      1042^{      +711}_{      -587}$ &     $      1238^{      +712}_{      -667}$ &     $       945^{      +655}_{      -525}$ &     $       905^{      +763}_{      -514}$ &     $      1281^{      +670}_{      -689}$      \\
$\Delta T$ between 32 and 10 bar         & K                    & $       300^{      +556}_{      -215}$ &     $      1122^{      +502}_{      -606}$ &     $       257^{      +555}_{      -181}$ &     $       539^{      +532}_{      -335}$ &     $       173^{      +164}_{      -103}$ &     $       302^{      +276}_{      -183}$ &     $      1015^{      +544}_{      -546}$      \\
$\Delta T$ between 10 and 3.2 bar        & K                    & $        68^{      +167}_{       -48}$ &     $       467^{      +448}_{      -274}$ &     $        61^{      +150}_{       -44}$ &     $       163^{      +299}_{      -117}$ &     $        45^{       +40}_{       -27}$ &     $        67^{       +57}_{       -40}$ &     $       666^{      +425}_{      -385}$      \\
$\Delta T$ between 3.2 and 1 bar         & K                    & $        37^{       +51}_{       -24}$ &     $        89^{      +103}_{       -54}$ &     $        42^{       +54}_{       -26}$ &     $        56^{       +99}_{       -40}$ &     $         8^{       +11}_{        -5}$ &     $        11^{       +14}_{        -7}$ &     $       265^{      +163}_{      -141}$      \\
$\Delta T$ between 1 and 0.1 bar         & K                    & $       112^{       +86}_{       -79}$ &     $        26^{       +24}_{       -16}$ &     $       142^{       +97}_{       -84}$ &     $        45^{       +48}_{       -27}$ &     $       306^{       +61}_{       -65}$ &     $        94^{       +39}_{       -49}$ &     $        86^{      +100}_{       -56}$      \\
$\Delta T$ between 0.1 bar and 1 mbar    & K                    & $       882^{       +66}_{       -80}$ &     $       562^{       +98}_{      -127}$ &     $       870^{       +67}_{       -75}$ &     $       797^{       +81}_{       -79}$ &     $       620^{       +65}_{       -59}$ &     $       638^{       +49}_{       -34}$ &     $       231^{       +52}_{       -62}$      \\
$\Delta T$ between 1 mbar and 10 nbar    & K                    & $       493^{      +290}_{      -280}$ &     $       759^{      +135}_{      -207}$ &     $       475^{      +299}_{      -269}$ &     $       591^{      +237}_{      -292}$ &     $       145^{      +111}_{       -74}$ &     $       220^{       +66}_{       -64}$ &     $       954^{       +30}_{       -54}$      \\
$\log$(mr$_{\rm{MgSiO}_3}$)              & \nodata              & $     -2.56^{     +0.56}_{     -0.55}$ &     $     -1.92^{     +0.36}_{     -0.31}$ &     $     -3.38^{     +0.37}_{     -0.39}$ &     $     -2.10^{     +0.54}_{     -0.61}$ &     $     -1.64^{     +0.34}_{     -0.36}$ &     $     -1.25^{     +0.14}_{     -0.25}$ &     $     -1.62^{     +0.35}_{     -0.58}$      \\
$\log$(K$_{zz}$)                         & cm$2\cdot$s$^{-1}$   & $      5.49^{     +0.29}_{     -0.22}$ &     $      6.10^{     +0.28}_{     -0.18}$ &     $      5.48^{     +0.21}_{     -0.16}$ &     $      5.56^{     +0.24}_{     -0.22}$ &     $      7.40^{     +0.26}_{     -0.21}$ &     $      8.42^{     +0.26}_{     -0.28}$ &     $      7.38^{     +0.19}_{     -0.24}$      \\
$f_{sed}$                                & \nodata              & $      0.63^{     +0.23}_{     -0.21}$ &     $      0.41^{     +0.15}_{     -0.11}$ &     $      0.44^{     +0.16}_{     -0.13}$ &     $      0.51^{     +0.20}_{     -0.20}$ &     $      0.31^{     +0.12}_{     -0.11}$ &     $      0.16^{     +0.06}_{     -0.04}$ &     $      0.36^{     +0.09}_{     -0.12}$      \\
Cloud log-normal size $\sigma_g$         & \nodata              & $      1.22^{     +0.19}_{     -0.10}$ &     $      1.29^{     +0.34}_{     -0.14}$ &     $      1.21^{     +0.20}_{     -0.09}$ &     $      1.26^{     +0.33}_{     -0.14}$ &     $      1.21^{     +0.15}_{     -0.09}$ &     $      1.22^{     +0.18}_{     -0.10}$ &     $      1.21^{     +0.36}_{     -0.11}$      \\
Extinction coefficient $\alpha$          & \nodata              & $      0.06^{     +0.05}_{     -0.04}$ &     $      0.05^{     +0.04}_{     -0.03}$ &     $      0.07^{     +0.05}_{     -0.04}$ &     $      0.07^{     +0.05}_{     -0.04}$ &     $      2.09^{     +0.22}_{     -0.14}$ &     $      2.24^{     +0.18}_{     -0.12}$ &     $      0.03^{     +0.03}_{     -0.02}$      \\
$\ln(EV)$ & & 0.00 & -23.76 & -1.47 & -15.86 & -304.19 & -365.29 & -20.43 \\
$\log(L/L_\odot)$ & & $-4.167^{+0.003}_{-0.002}$ & $-4.160^{+0.003}_{-0.003}$ & $-4.168^{+0.003}_{-0.002}$ & $-4.164^{+0.002}_{-0.003}$ & $-4.160^{+0.003}_{-0.003}$ & $-4.158^{+0.004}_{-0.002}$ & $-4.158^{+0.003}_{-0.002}$ \\
T$_{\rm{eff}}$ & K & $1477^{+8}_{-10}$ & $1480^{+12}_{-11}$ & $1463^{+10}_{-7}$ & $1464^{+10}_{-12}$ & $1515^{+14}_{-15}$ & $1540^{+9}_{-15}$ & $1511^{+7}_{-8}$ \\
\enddata


\end{deluxetable}

\end{CJK*}
 
\end{document}